\documentclass[aps,prl,twocolumn,showpacs,superscriptaddress,reprint]{revtex4-1}
\usepackage{graphics}
\usepackage{xcolor}   
\usepackage{amsmath}
\usepackage{epsfig}
\usepackage[utf8]{inputenc}

\begin{document}
\title{Effects of high order interatomic potential on elastic phonon scatterings}
\author{Hangbo Zhou}
\affiliation{Institute of High Performance Computing, A*STAR, 138632, Singapore}
\author{Gang Zhang}
\affiliation{Institute of High Performance Computing, A*STAR, 138632, Singapore}
\author{Yong-Wei Zhang}
\affiliation{Institute of High Performance Computing, A*STAR, 138632, Singapore}
\date{\today}
\begin{abstract}
Interatomic potentials beyond quadratic order provide scattering sources for phonon transport in lattice. By using a weakly-interacting interface model, we investigated the relation between the order of interatomic potential and the multiple-phonon scattering process. We find that high order interatomic potential not only causes multiple-phonon scattering processes, but also has significant impacts on elastic phonon scattering processes. Using fourth order potential as an example, we show that it can significantly affects elastic phonon scatterings, through the formation of localized phonons. Such impact is closely related to the correlations of interfacial atoms and it becomes more significant with increasing temperature. Our work suggests that it is insufficient to consider only quadratic potential to investigate elastic phonon transport.
\end{abstract}
\maketitle

\section{Introduction}
In many materials, phonon transport is responsible for heat conduction \cite{Li2012, Chen2000}. However, along the path of travelling, phonons will experience various scatterings due to various reasons, which can significantly influence the heat conductance. 
Such scatterings include, for example, elastic scatterings \cite{Zhao2005, Kothari2019} due to the mismatches of phonon spectral density and inelastic scatterings \cite{He2008,Fereidani2019,Reid2019, Guo2020,Cammarata2021} due to the anharmonic lattice vibrations. 
Across an interface, heat conductance across an interfaces is traditionally modelled phenomenologically by acoustic mismatch model (AMM) and diffusion mismatch model (DMM) . They provide two extreme cases that heat are either completely carried by elastic waves (AMM) or diffusive phonons (DMM), respectively. However, phonons will experience both elastic and inelastic scatterings \cite{Hopkins2009,Hopkins2009b}.

Rigorous developement of quantum theory of phonon transport based on atomistic model has been established in the last two decades using Non-equilibrium Green's function (NEGF) technique \cite{Wang2006, Wang2008}.
NEGF predicts that for a harmonic lattice (the interatomic potential is quadratic with respect to vibrational displacement), phonon experience only elastic scatterings, which means the phonon energy is conserved . 
Since then, many studies of elastic phonon scatterings have been reported, applied to many materials or nanostructures \cite{Jiang2009,  Ouyang2010, Zhou2016,Ju2017, Zhou2017, Zhou2018}. In this approach, the lattice potential is approximated by quadratic potential and this assumption is justified at low temperature. 
Often, Elastic phonon transport also serves as a foundation to understand more-involved multiple-phonon scattering process \cite{Fereidani2019}. 

%By introducing an interface, elastic phonon transport can be described as elastic scattering process, in the sense that the transport process across the interface is through absorbing a phonon from one side, and emitting the other phonon with same energy to the other side.

%This assumption holds in the condition of low temperature and strong interatomic bonds such that the vibrational displacements is small. For potentials up to quadratic order of displacements, it is exactly solvable with modern theories such non-equilibrium Green's function formalism . Numerous works has been done under this framework for heat transport in low-dimensional nanostructures with fruitful findings. 

So far the the analysis of elastic scattering is still limited to interatomic potentials within quadratic order. For potentials beyond quadratic order, it turns out to be extremely challenging to solve exactly and quantum mechanically \cite{Xu2008, Guo2020}. As a result, the relation between higher order potential and multiple-phonon scattering processes is much less understood. From the Fermi' Golden rule we can understand that if the order of potential reaches $n$, the maximum phonons involved in the scattering is $n$. For instance, a cubic order potential is able to cause three-phonon process and a fourth order potential is able to cause four-phonon processes \cite{Feng2016,Feng2017,Ravichandran2020}. 
However, whether a high order potential has impacts on elastic phonon scatterings has not been addressed. 

In this work we will investigate the relation between higher order potentials and elastic phonon scattering process. In order to bypass the difficulties of exactly solving high order potential problems, we introduce an anharmonic interface and limit the coupling of the interface to be weak, so that it can be treated perturbatively. Such perturbation treatment will not obscure the rendering of phonon scatterings and thus it provides a opportunity to discover the role higher order potentials to phonon scattering process. 
Furthermore, heat transfer such weakly interacting interface has important applications such as the in-plane heat conductance through van der Waals heterostructures \cite{Tielrooij2018, Alborzi2020}.
To our surprise, we find that a $n$-th order potential will not only cause $n$-phonon process, but also takes important roles to elastic phonon scattering processes. Depending on the details of potential, it can either enhance or suppress the elastic scatterings.

\section{Theoretical derivation}

We model our interface by connecting two harmonic thermal baths via interfacial couplings.  In general, the Hamiltonian can be written as
\begin{equation}
    H=H_L+H_R+H_{int},
\end{equation}
where $H_L=\sum_q\frac{(\tilde{p}^L_q)^2}{2m}+\frac{1}{2}\omega_q^2(\tilde{x}_q^L)^2$ and $H_R=\sum_q\frac{(\tilde{p}^R_q)^2}{2m}+\frac{1}{2}\omega_q^2(\tilde{x}_q^R)^2$ are collections of harmonic oscillators. The interfacial couplings consist of both quadratic couplings and higher order couplings. To be specific, we used the potential up to fourth order of interatomic forces,
\begin{eqnarray}
    &V&=\frac{1}{2!}\sum_{ij}K_{i,j}x_ix_j+\frac{1}{3!}\sum_{ijk}(V_{ij,k}x_ix_jx_k+V_{i,jk}x_ix_jx_k)\nonumber\\
    &+&\frac{1}{4!}\sum_{ijkl}(T_{ijk,l}x_ix_jx_kx_l+T_{ij,kl}x_ix_jx_kx_l+T_{i,jkl}x_ix_jx_kx_l),\nonumber
\end{eqnarray}
where $K_{i,j}$ are interatomic force constants (IFCs) of quadratic coupling, $V$ are the IFCs of cubic couplings and $T$ are the IFCs of the fourth order couplings. The displacement, for example $x^L_i$, can be expanded with respect to the displacement of the normal modes of phonons with wave vector $q$ in $L$ as $x^L_i=\sum_q c_i^q \tilde{x}^L_q$. 

For the calculations of thermal current we employ the formalism developed in ref \cite{Zhou2020}. In the weak interaction regime, The thermal current is determined by the correlations of the operators that is involved in the interface coupling. It is well-understood that the quadratic coupling between the two baths only causes elastic scattering processes. In other words, the phonons are transmitted through without changing their energies. For the elastic scatterings caused by quadratic coupling, its contribution to thermal conductance can be written as \cite{Zhou2020}
\begin{equation}
\label{eq:2ph}
    I_{2p}=-\frac{1}{4\hbar}\sum_{ij,kl}K_{i,j}K_{k,l}\int_{-\infty}^\infty\Psi_{ik}(t)\Phi_{jl}(t)dt,
\end{equation}
where $\Psi_{ij}(t)=\frac{d\Phi_{ij}(t)}{dt}$, $\Phi_{ij}(t)=\left<x_i(t)x_j\right>$ are the two-point displacement correlation functions. 
The two-point correlation functions can be written in terms of the spectral densities of the left part $\Gamma_L$ and right part $\Gamma_R$ as $\Phi_{ij}(t)=\int_{-\infty}^\infty\frac{d\omega}{\pi}\Gamma_{ij}(\omega)n(\omega)e^{i\omega t}$ and $\Psi_{ij}(t)=i\int_{-\infty}^\infty\frac{d\omega}{\pi}\Gamma_{ij}(\omega)\omega n(\omega)e^{i\omega t}$. A straightforward derivation will show that it can be cast into Landauer formula,
\begin{equation}
    I_{2p}=\frac{1}{(2!)^2}\sum_{ijkl}K_{i,j}K_{k,l} H_{ikjl},
\end{equation}
where
\begin{equation}
    H_{ikjl}=\frac{4}{\hbar}\int_0^\infty \frac{d\omega}{\pi}\omega J^L_{ik}(\omega)J^R_{jl}(-\omega)[n^L(\omega)-n^R(\omega)].
\end{equation}

\begin{figure}
    \centering
    \includegraphics[width=\linewidth]{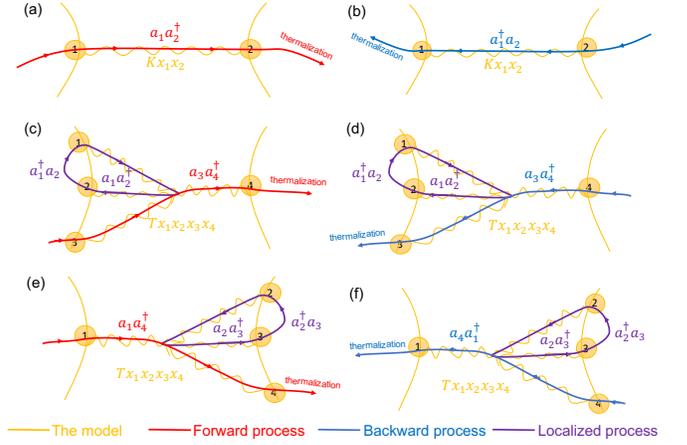}
    \caption{Phenomenological illustration of elastic phonon scattering caused by atoms that involves quadratic couplings (a and b) and fourth order couplings (c-f).  The origin symbols shows the model, the red lines show the forward scattering process, the blue lines show the backward scattering process and the brown lines shows localized processes that does not involve energy transmission from left side to right side.}
    \label{fig:ill}
\end{figure}

This results can also be derived from NEGF approach with bath-bath coupling developed in the literature\cite{Li2012a}. Such elastic scattering processes are phenomenological shown in Fig.~\ref{fig:ill}(a) and Fig.~\ref{fig:ill}(b). In the figure we focus on a phonon with specific frequency $\omega$, since the phonon energy and phonon frequency will not change during the elastic scattering. The atoms labelled with 1 and 2 are the interfacial atoms that involve interactions with the atoms of the other side of interface. The coupling between atom 1 and 2 are in quadratic order. Through the coupling, a phonon of energy ($\hbar\omega$) can be annihilated at atom 1 and simultaneously a phonon of same energy can be created at atom 2. through this process, a energy of amount $\hbar\omega$ is transmitted across the interface. Upon transmission, this phonon will be dissipated  and thermalized into the thermal bath of the right side. As required by detailed balance, phonons will also enter a backward scattering process shown in blue in Fig.~\ref{fig:ill}(b), where a phonon of energy $\hbar\omega$ is transmitted from atom 2 to 1. During the cycles of phonon creation and annihilation between atom 1 and 2, phonons are either dissipated into or emitted from baths of both sides. The mount of net heat flow is determined by the competition between the phonon emission and dissipation rates at both baths. For example, if the left bath has larger emission ratio (Fig.~\ref{fig:ill}(a)) than dissipation ratio (Fig.~\ref{fig:ill}(b)), then the right bath should have larger dissipation ratio than emission ratio. As a result, heat will flow from the left side to the right side. However, the magnitude of heat conductance will be determined by the amount of energy carried by the phonon and the occurrence probability of the scattering processes. So in detail, they are determined the phonon spectra density, the phonon occupation number (temperature) and the strength of the quadratic coupling. Mathematically they are summarized in Eq.~(\ref{eq:2ph}).

For the third order coupling, we have shown that it contributes to three-phonon processes, which consists of phonon splitting, merging, partial reflection and partial transmission \cite{Zhou2020}. However, in the weak coupling regime, its existence will not affect the elastic scattering processes. In other words, the quadratic coupling contributes elastic scatterings and cubic coupling contributes to three-phonon scatterings. Their effects on thermal conductance are separable are additive. 

The interesting roles come from the fourth order coupling at the interface. Firstly, we find that, the fourth order coupling do contribute to the four-phonon processes as expected, which involve processes of a single phonon splitting into three phonons, or three phonons merging into one, or two phonons merging together with emission of two new phonons. However, surprisingly, in addition to the four-phonon processes, the fourth order coupling also affects the elastic scattering processes. 

Mathematically, such effects come from both the cross term between quadratic and fourth order coupling, and the fourth order couplings alone. In the following, we will analyze it in detail. 

We first analyze the contribution of the correlation between term $\sum_{ij}K_{i,j}x_ix_j$ and term $\frac{1}{4!}\sum_{ijkl}T_{ijk,l}x_ix_jx_kx_l$, where the forward process is mediated via the coupling of $\sum_{ij}K_{i,j}x_ix_j$ while the backward process is mediated via the coupling $\frac{1}{4!}\sum_{ijkl}T_{ijk,l}x_ix_jx_kx_l$.
The evaluation of this term involve the calculation of four-point correlation function. By using Wick's theorem we can find that 
\begin{equation}
    \phi_{ijkl}^L(t)=\langle x_i(t)x_jx_kx_l\rangle=c^L_{ij}(t)Z^L_{kl}+c^L_{ik}(t)Z^L_{jl}+c^L_{il}(t)Z^L_{jk}.\nonumber
\end{equation}
Here we have defined $Z_{ij}=c_{ij}(t=0)$ as the correlation function at equal time. If $i=j$, it is the expectation value of square of the amplitude of atomic vibration. Therefore, it increases with temperature as well as the spectral density of that atom. In the high temperature limit, it should be proportional to temperature according to equal partition theorem. 

With the correlation function, we find that its contribution to thermal current is 
\begin{equation}
I_h=\frac{1}{2!}\frac{1}{4!}\sum_{ijklmn}6K_{i,j}T_{klm,n}Z^L_{lm}H_{ikjn}
\end{equation}
We immediately find that its contribution to thermal current is proportional to $Z^L$, which provides a extra temperature-dependent components. As we know $H_{ikjn}$ will saturate in the high temperature limit. So this term will eventually be linearly increasing with temperature of left part.

This phonon scattering process can be phenomenological explained through the scattering process shown in Fig.~\ref{fig:ill}. It describes a combination of two cycles of scattering processes. 
In the first cycle,  the forward process is carried via the quadratic coupling Fig.~\ref{fig:ill}(a), the backward scattering is through the fourth order coupling Fig.~\ref{fig:ill}(d). In this backward scattering process, the transmitted phonon maintain the same energy across the interface. So it is regarded as elastic scattering.  It happens when the other two atoms that involved in the fourth order coupling, atom 1 and 2, forms a localized phonon mode, such that the phonon forms a closed cycle in the left bath, and it is not travelling to the other side of interface. 
However, whenever a elastic scattering from 4 to 3 is happened,  a phonon conversion is simultaneously occurred between atom 1 and 2, due to the fact that their interatomic coupling is in fourth order. In such a way, the localized phonons between atom 1 and 2 will significantly affects the scattering probability between 3 and 4, and such affects the total heat conduction. Such effects is quantitatively described by the quantity $Z_L$.  

In the other cycle, on the contrast, the forward process is carried by the fourth order coupling Fig.~\ref{fig:ill}(c) and the backward process is carried by the quadratic coupling Fig.~\ref{fig:ill}(b). Similarly, the direction of net heat flow is determined by the phonon emission and dissipation ratio of the two baths. 

In the above, we have shown a typical example that a high order potential can cause elastic phonon scattering, through the formation of localized phonons. 
In a similar manner, we can also calculate the other contributions. Next we consider the cross term between $\sum_{ij}K_{i,j}x_ix_j$ and $\frac{1}{4!}\sum_{ijkl}T_{i,jkl}x_ix_jx_kx_l$. It turns out to be
\begin{equation}
   I_h=\frac{1}{2!}\frac{1}{4!}\sum_{ijklmn}3K_{i,j}T_{k,lmn}Z^R_{mn}H_{ikjl} 
\end{equation}
Its phenomenological illustration can be described by two cycles. One of which is illustrated by \ref{fig:ill}(a) and Fig.~\ref{fig:ill}(f), and the other is illustrated by Fig.~\ref{fig:ill}(b) Fig.~\ref{fig:ill}(e). We can find that this term depends on the $Z^R$. This term and previous term together cause asymmetry between the left and right bath and thus they will result in thermal rectification effect under temperature bias. 

For the cross term between $\sum_{ij}K_{i,j}x_ix_j$ and $\sum_{mnop}T_{mn,op,kl}x_mx_nx_ox_p$, it will not contribute to thermal current, because a closed cycle is not able to be formed. 

So far we have analyze the cross term between the quadratic and fourth order coupling. We show that the cross term will increase linearly with temperature in the high temperature limit. Next we will show the terms coming from solely the fourth order coupling. We first analyze the term from coupling between $T_{ijk,l}$ and $T_{mno,p}$. We find that such coupling not only contributes to four-phonon process, but also to elastic scattering process. Its contribution on elastic scattering process is illustrated through cycles formed by Fig.~\ref{fig:ill}(c) and Fig.~\ref{fig:ill}(d). In this case, two localized phonons are formed at left size, which affecting both the forward and backward processes. Mathematically, its contribution to thermal current is given by
\begin{equation}
    I_{h(2p)}=\frac{1}{(4!)^2}\sum_{ijklmnop}9T_{ijk,l}T_{mno,p}Z^L_{ij}Z^L_{no}H_{kmlp}
\end{equation}
We find that it depends on the second order of $Z$ and increase quadratically in the high temperature limit. Similarly, the contribution from coupling between $T_{l,ijk}$ and $T_{p,mno}$ is
\begin{equation}
     I_{h(2p)}=\frac{1}{(4!)^2}\sum_{ijklmnop}9T_{l,ijk}T_{p,mno}Z^R_{ij}Z^R_{no}H_{lpkm}
\end{equation}
and it is illustrated by Fig.~\ref{fig:ill}(e) and Fig.~\ref{fig:ill}(f). 
The contribution from coupling between $T_{ijk,l}$ and $T_{m,nop}$ is
\begin{equation}
  I_{h(2p)}=\frac{1}{(4!)^2}\sum_{ijklmnop}9T_{ijk,l}T_{m,nop}Z^L_{ij}Z^R_{op}H_{kmln}  
\end{equation}
and it is illustrated in Fig.~\ref{fig:ill}(f).
By defining \begin{equation}
    S^L_{ij}=\sum_{mn\in L}T_{imn,j}Z^L_{mn}, \,\,\,\,\,S^R_{ij}=\sum_{mn\in R}T_{i,mnj}Z^R_{mn}
\end{equation}
%\begin{equation}
%    S^R_{ij}=\sum_{mn\in R}T_{i,mnj}Z^R_{mn}
%\end{equation}
We also find that cross term between $\sum_{ijkl}T_{ij,jk}x_ix_jx_kx_l$ and $\sum_{mnop}T_{mn,op,kl}x_mx_nx_ox_p$ will not contribute to the elastic scatterings. In this case, only the four-phonon processes contribute to the phonon transport. The localized phonons, even formed, are not able to affect the phonon transport processes.

By summerizing all the contributions, we find that the total elastic scattering processes can be written in a concise form as
\begin{eqnarray}
    I_{h(2p)}&=&\sum_{ij,kl}H_{ijkl}\nonumber\\
    &\times&\Big(\frac{1}{2!}K_{ik}+\frac{3}{4!}(S^L_{ik}+S^R_{ik})\Big)\Big(\frac{1}{2!}K_{jl}+\frac{3}{4!}(S^L_{jl}+S^R_{jl})\Big)\nonumber\\
\end{eqnarray}

Hence we have shown that the fourth order coupling at interface will have an impact on elastic scattering process. We found a temperature dependence quantity $S$ that can be regarded as a effective quadratic force constant. The value of $S$ will increase linearly with temperature. Therefore, such impacts will increase with increasing of temperature and it will eventually dominate at high temperature regime. This formalism suggest that even in the evaluation of elastic scattering process, it is insufficient to consider only the quadratic interatomic force constant $K$. One need to evaluate the effective force constant $S$ from fourth order potential. With even higher temperature, there should be even contributions from higher order potentials.
\section{Numerical results of an application}

\begin{figure}
    \centering
    \includegraphics[width=\linewidth]{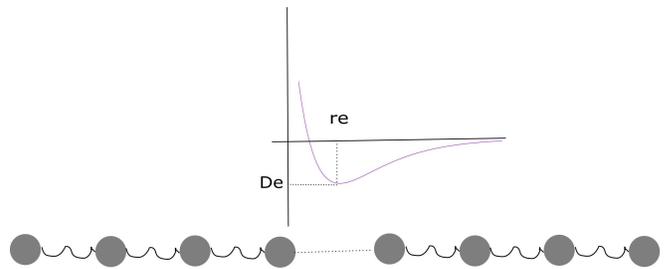}
    \caption{Illustration of the setup used in our calculation. The interface is modelled by Morse potential, which has minimum energy of $D_e$ at equilibrium position $r_e$. The left and right lead are Rubin baths. 
    }
    \label{fig:potential}
\end{figure}

Next we use one-dimension chain to demonstrate such effects. In the simplest model, the interface is comprised by connecting to Rubin baths as shown in Fig.~\ref{fig:potential}. At the interface, only the nearest atoms are interacting with each other. The interatomic force constant within the bath are characterized by $k$. At the interface, we use Morse potential $V(r)=De(e^{-2a(r-r_e)}-2e^{-a(r-r_e)})$ to simulate the coupling potential between two baths. The interatomic potential within the lead $K$ is significantly larger interface coupling.

In our setup, we allow to add external forces to adjust the interatomic distance at the interface, and hence will adjust the interatomic potential. If we stretch the two leads, then both the interatomic distance between the leads and within the leads will increase. They will reach a new equilibrium position at the point where $V'(r)=kr$. The interfacial atoms at this new position is balanced by both the Morse potential and the quadratic potential within the lead. However, with regarding to the interface coupling, both the second order and higher order force constants are adjusted. The quadratic and fourth-order IFCs can be calculated via derivatives of Morse potential with respect that new equilibrium position. 

\begin{figure}
    \centering
    \includegraphics[width=0.8\linewidth]{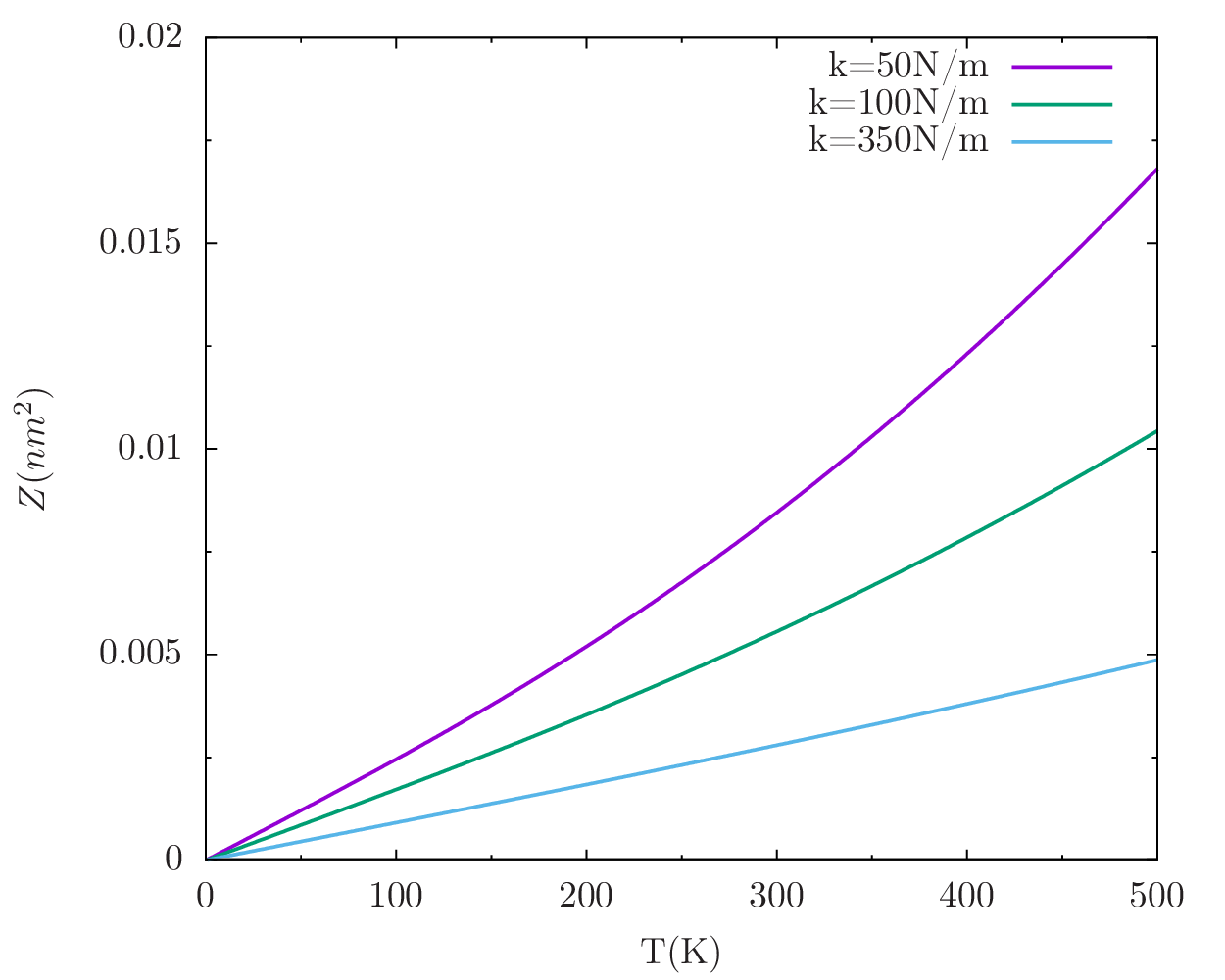}
    \caption{The mean square of vibrational amplitude $Z$ is plotted against temperature under different interatomic force constants $k=50N/m$, $k=100N/m$ and $k=350N/m$. }
    \label{fig:ZT}
\end{figure}

Fig.~\ref{fig:ZT} shows the temperature dependence of $Z$, which can be regarded as the mean square of vibrational amplitude of interfacial atoms. It shows $Z$ increase with increase of temperature. Theoretically it will eventually linear increase with respect to temperature in the high temperature regime. The figure also shows $Z$ is larger when $k$ is smaller. The magnitude of $Z$ in comparison with the ratio of fourth order and second order IFCs ($\eta=T/K$) will determine how important will the fourth order potential be on elastic scattering processes. If $Z$ is comparable to $\eta$, the impact of 4th order IFCs will have comparable effects on elastic scattering processes with respect to quadratic IFCs. As a result, we can conclude that 1) 4th order potential is less important in low temperature regime and increasingly important with increase of temperature. This result is consistent with previous findings in the literature. \cite{Reid2019} 2) 4th order potential is more important when the bonds in the baths are weaker (smaller $k$) but less important for stronger bonds. This is consistent with literature that elastic scatterings normally dominates for phonons in graphene, which has strong carbon-carbon bonds \cite{Balandin2020}.

\begin{figure}
    \centering
    \includegraphics[width=\linewidth]{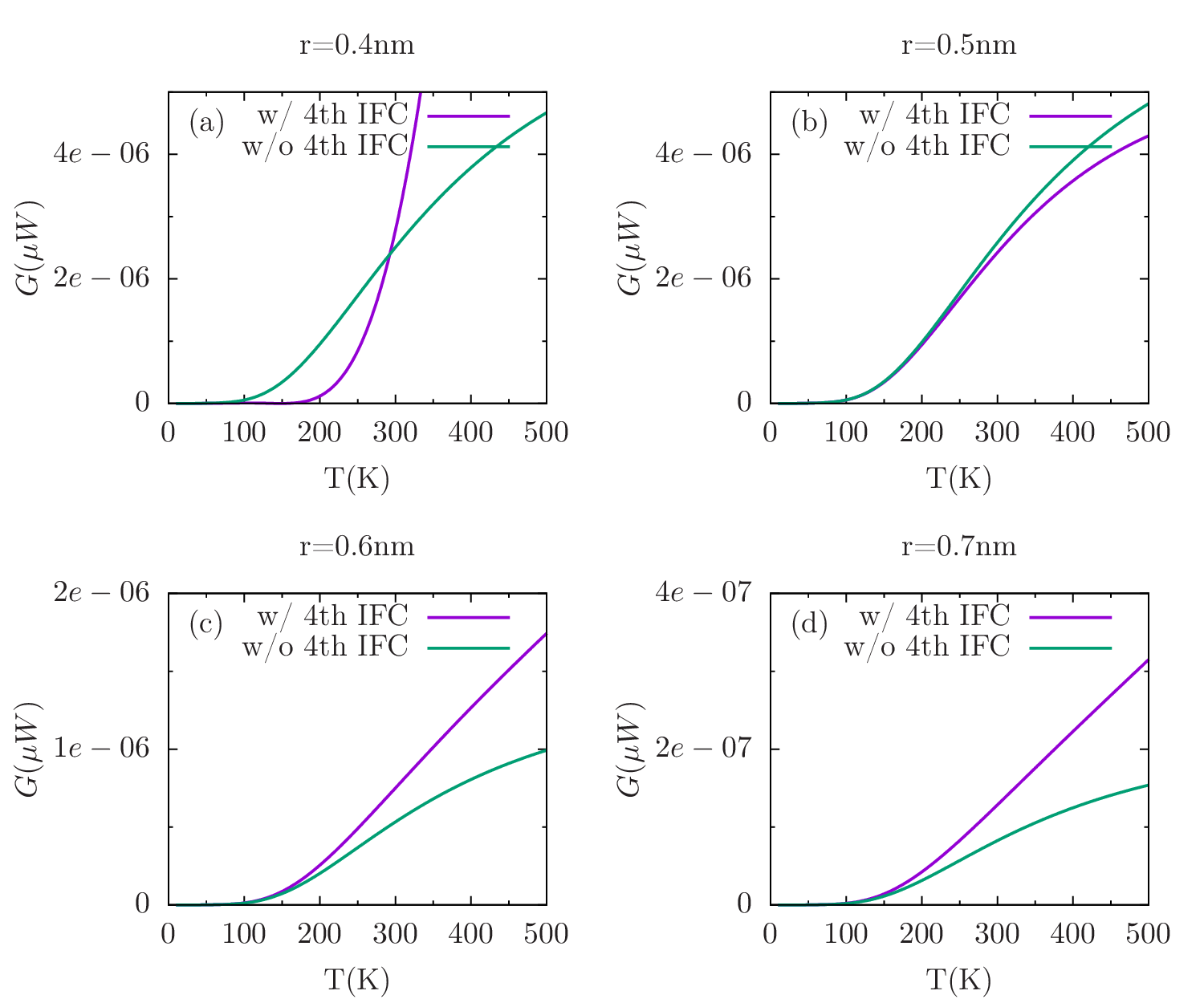}
    \caption{
    Temperature dependence of thermal conductance with or without 4th order coupling under different distances. Parameters: a=1/A. De=0.5eV. k=350N/m. re=0.3nm.}
    \label{fig:result}
\end{figure}

Fig.~\ref{fig:result} shows the temperature dependence of the contribution of elastic scattering processes to the thermal conductance, with and without considering of 4th order IFCs. We used the interatomic distance at interface to adjust the potential.  We find that 4th order IFC can both enhance or suppress the elastic scattering process at different distance and temperatures. When $r=0.4nm$, the 4th order potential suppress elastic scattering process at low temperature but enhance it at high temperature with crossover at around $T=300K$. When $r=0.5nm$, 4th order IFCs will suppress the elastic scattering process while at $t=0.6nm$ and $r=0.7nm$ it will enhance the elastic scattering processes. In this particular one-dimensional case, the effects is determined by the sign of the 2nd and 4th order IFC. Specifically, when $r=0.5$nm, The 4th order potential will suppress elastic scattering process when the sign of 4th IFC is different from that of 2nd order IFC. For $r=$ 0.4nm, 0.6nm and 0.7nm, The 4th order potential will enhance the elastic scattering elastic scattering since the sign of 4th IFC is the same as that of 2nd IFC

\section{Conclusion}
During heat conduction in lattice, a quadratic potential can cause elastic scatterings for the travelling phonons. It will manifest as elastic scattering processes at interface. Higher order nonlinear potentials will responsible for multiple phonon scattering processes. In this work, we find that the 4th order potential has significant effect on elastic scattering process as well. This effect will be more significant with increase of temperature. From our model calculation, it shows that the 4th order potential can either enhance or suppress the elastic scattering process, depending to the coupling coefficients as well as temperature regime. This work suggests that in order to completely evaluate the elastic scattering process of phonon transport in lattice, one need to consider quadratic potential as well as higher order potentials, especially when the temperature is not sufficiently low.

\bibliography{four}

\begin{thebibliography}{28}%
\makeatletter
\providecommand \@ifxundefined [1]{%
 \@ifx{#1\undefined}
}%
\providecommand \@ifnum [1]{%
 \ifnum #1\expandafter \@firstoftwo
 \else \expandafter \@secondoftwo
 \fi
}%
\providecommand \@ifx [1]{%
 \ifx #1\expandafter \@firstoftwo
 \else \expandafter \@secondoftwo
 \fi
}%
\providecommand \natexlab [1]{#1}%
\providecommand \enquote  [1]{``#1''}%
\providecommand \bibnamefont  [1]{#1}%
\providecommand \bibfnamefont [1]{#1}%
\providecommand \citenamefont [1]{#1}%
\providecommand \href@noop [0]{\@secondoftwo}%
\providecommand \href [0]{\begingroup \@sanitize@url \@href}%
\providecommand \@href[1]{\@@startlink{#1}\@@href}%
\providecommand \@@href[1]{\endgroup#1\@@endlink}%
\providecommand \@sanitize@url [0]{\catcode `\\12\catcode `\$12\catcode
  `\&12\catcode `\#12\catcode `\^12\catcode `\_12\catcode `\%12\relax}%
\providecommand \@@startlink[1]{}%
\providecommand \@@endlink[0]{}%
\providecommand \url  [0]{\begingroup\@sanitize@url \@url }%
\providecommand \@url [1]{\endgroup\@href {#1}{\urlprefix }}%
\providecommand \urlprefix  [0]{URL }%
\providecommand \Eprint [0]{\href }%
\providecommand \doibase [0]{http://dx.doi.org/}%
\providecommand \selectlanguage [0]{\@gobble}%
\providecommand \bibinfo  [0]{\@secondoftwo}%
\providecommand \bibfield  [0]{\@secondoftwo}%
\providecommand \translation [1]{[#1]}%
\providecommand \BibitemOpen [0]{}%
\providecommand \bibitemStop [0]{}%
\providecommand \bibitemNoStop [0]{.\EOS\space}%
\providecommand \EOS [0]{\spacefactor3000\relax}%
\providecommand \BibitemShut  [1]{\csname bibitem#1\endcsname}%
\let\auto@bib@innerbib\@empty
%</preamble>
\bibitem [{\citenamefont {Li}\ \emph {et~al.}(2012{\natexlab{a}})\citenamefont
  {Li}, \citenamefont {Ren}, \citenamefont {Wang}, \citenamefont {Zhang},
  \citenamefont {H\"anggi},\ and\ \citenamefont {Li}}]{Li2012}%
  \BibitemOpen
  \bibfield  {author} {\bibinfo {author} {\bibfnamefont {N.}~\bibnamefont
  {Li}}, \bibinfo {author} {\bibfnamefont {J.}~\bibnamefont {Ren}}, \bibinfo
  {author} {\bibfnamefont {L.}~\bibnamefont {Wang}}, \bibinfo {author}
  {\bibfnamefont {G.}~\bibnamefont {Zhang}}, \bibinfo {author} {\bibfnamefont
  {P.}~\bibnamefont {H\"anggi}}, \ and\ \bibinfo {author} {\bibfnamefont
  {B.}~\bibnamefont {Li}},\ }\href {\doibase 10.1103/RevModPhys.84.1045}
  {\bibfield  {journal} {\bibinfo  {journal} {Rev. Mod. Phys.}\ }\textbf
  {\bibinfo {volume} {84}},\ \bibinfo {pages} {1045} (\bibinfo {year}
  {2012}{\natexlab{a}})}\BibitemShut {NoStop}%
\bibitem [{\citenamefont {Chen}(2000)}]{Chen2000}%
  \BibitemOpen
  \bibfield  {author} {\bibinfo {author} {\bibfnamefont {G.}~\bibnamefont
  {Chen}},\ }\href {\doibase https://doi.org/10.1016/S1290-0729(00)00202-7}
  {\bibfield  {journal} {\bibinfo  {journal} {International Journal of Thermal
  Sciences}\ }\textbf {\bibinfo {volume} {39}},\ \bibinfo {pages} {471}
  (\bibinfo {year} {2000})}\BibitemShut {NoStop}%
\bibitem [{\citenamefont {Zhao}\ and\ \citenamefont {Freund}(2005)}]{Zhao2005}%
  \BibitemOpen
  \bibfield  {author} {\bibinfo {author} {\bibfnamefont {H.}~\bibnamefont
  {Zhao}}\ and\ \bibinfo {author} {\bibfnamefont {J.~B.}\ \bibnamefont
  {Freund}},\ }\href {\doibase 10.1063/1.1835565} {\bibfield  {journal}
  {\bibinfo  {journal} {Journal of Applied Physics}\ }\textbf {\bibinfo
  {volume} {97}},\ \bibinfo {pages} {024903} (\bibinfo {year}
  {2005})}\BibitemShut {NoStop}%
\bibitem [{\citenamefont {Kothari}\ \emph {et~al.}(2019)\citenamefont
  {Kothari}, \citenamefont {Malhotra},\ and\ \citenamefont
  {Maldovan}}]{Kothari2019}%
  \BibitemOpen
  \bibfield  {author} {\bibinfo {author} {\bibfnamefont {K.}~\bibnamefont
  {Kothari}}, \bibinfo {author} {\bibfnamefont {A.}~\bibnamefont {Malhotra}}, \
  and\ \bibinfo {author} {\bibfnamefont {M.}~\bibnamefont {Maldovan}},\ }\href
  {\doibase 10.1088/1361-648x/ab2172} {\bibfield  {journal} {\bibinfo
  {journal} {Journal of Physics: Condensed Matter}\ }\textbf {\bibinfo {volume}
  {31}},\ \bibinfo {pages} {345301} (\bibinfo {year} {2019})}\BibitemShut
  {NoStop}%
\bibitem [{\citenamefont {He}\ \emph {et~al.}(2008)\citenamefont {He},
  \citenamefont {Buyukdagli},\ and\ \citenamefont {Hu}}]{He2008}%
  \BibitemOpen
  \bibfield  {author} {\bibinfo {author} {\bibfnamefont {D.}~\bibnamefont
  {He}}, \bibinfo {author} {\bibfnamefont {S.}~\bibnamefont {Buyukdagli}}, \
  and\ \bibinfo {author} {\bibfnamefont {B.}~\bibnamefont {Hu}},\ }\href
  {\doibase 10.1103/PhysRevE.78.061103} {\bibfield  {journal} {\bibinfo
  {journal} {Phys. Rev. E}\ }\textbf {\bibinfo {volume} {78}},\ \bibinfo
  {pages} {061103} (\bibinfo {year} {2008})}\BibitemShut {NoStop}%
\bibitem [{\citenamefont {Moghaddasi~Fereidani}\ and\ \citenamefont
  {Segal}(2019)}]{Fereidani2019}%
  \BibitemOpen
  \bibfield  {author} {\bibinfo {author} {\bibfnamefont {R.}~\bibnamefont
  {Moghaddasi~Fereidani}}\ and\ \bibinfo {author} {\bibfnamefont
  {D.}~\bibnamefont {Segal}},\ }\href {\doibase 10.1063/1.5075620} {\bibfield
  {journal} {\bibinfo  {journal} {The Journal of Chemical Physics}\ }\textbf
  {\bibinfo {volume} {150}},\ \bibinfo {pages} {024105} (\bibinfo {year}
  {2019})}\BibitemShut {NoStop}%
\bibitem [{\citenamefont {Reid}\ \emph {et~al.}(2019)\citenamefont {Reid},
  \citenamefont {Pandey},\ and\ \citenamefont {Leitner}}]{Reid2019}%
  \BibitemOpen
  \bibfield  {author} {\bibinfo {author} {\bibfnamefont {K.~M.}\ \bibnamefont
  {Reid}}, \bibinfo {author} {\bibfnamefont {H.~D.}\ \bibnamefont {Pandey}}, \
  and\ \bibinfo {author} {\bibfnamefont {D.~M.}\ \bibnamefont {Leitner}},\
  }\href {\doibase 10.1021/acs.jpcc.8b11640} {\bibfield  {journal} {\bibinfo
  {journal} {The Journal of Physical Chemistry C}\ }\textbf {\bibinfo {volume}
  {123}},\ \bibinfo {pages} {6256} (\bibinfo {year} {2019})}\BibitemShut
  {NoStop}%
\bibitem [{\citenamefont {Guo}\ \emph {et~al.}(2020)\citenamefont {Guo},
  \citenamefont {Bescond}, \citenamefont {Zhang}, \citenamefont {Luisier},
  \citenamefont {Nomura},\ and\ \citenamefont {Volz}}]{Guo2020}%
  \BibitemOpen
  \bibfield  {author} {\bibinfo {author} {\bibfnamefont {Y.}~\bibnamefont
  {Guo}}, \bibinfo {author} {\bibfnamefont {M.}~\bibnamefont {Bescond}},
  \bibinfo {author} {\bibfnamefont {Z.}~\bibnamefont {Zhang}}, \bibinfo
  {author} {\bibfnamefont {M.}~\bibnamefont {Luisier}}, \bibinfo {author}
  {\bibfnamefont {M.}~\bibnamefont {Nomura}}, \ and\ \bibinfo {author}
  {\bibfnamefont {S.}~\bibnamefont {Volz}},\ }\href {\doibase
  10.1103/PhysRevB.102.195412} {\bibfield  {journal} {\bibinfo  {journal}
  {Phys. Rev. B}\ }\textbf {\bibinfo {volume} {102}},\ \bibinfo {pages}
  {195412} (\bibinfo {year} {2020})}\BibitemShut {NoStop}%
\bibitem [{\citenamefont {Cammarata}\ and\ \citenamefont
  {Polcar}(2021)}]{Cammarata2021}%
  \BibitemOpen
  \bibfield  {author} {\bibinfo {author} {\bibfnamefont {A.}~\bibnamefont
  {Cammarata}}\ and\ \bibinfo {author} {\bibfnamefont {T.}~\bibnamefont
  {Polcar}},\ }\href {\doibase 10.1103/PhysRevB.103.035406} {\bibfield
  {journal} {\bibinfo  {journal} {Phys. Rev. B}\ }\textbf {\bibinfo {volume}
  {103}},\ \bibinfo {pages} {035406} (\bibinfo {year} {2021})}\BibitemShut
  {NoStop}%
\bibitem [{\citenamefont {Hopkins}(2009)}]{Hopkins2009}%
  \BibitemOpen
  \bibfield  {author} {\bibinfo {author} {\bibfnamefont {P.~E.}\ \bibnamefont
  {Hopkins}},\ }\href@noop {} {\bibfield  {journal} {\bibinfo  {journal}
  {Journal of Applied Physics}\ }\textbf {\bibinfo {volume} {106}},\ \bibinfo
  {pages} {013528} (\bibinfo {year} {2009})}\BibitemShut {NoStop}%
\bibitem [{\citenamefont {Hopkins}\ and\ \citenamefont
  {Norris}(2009)}]{Hopkins2009b}%
  \BibitemOpen
  \bibfield  {author} {\bibinfo {author} {\bibfnamefont {P.~E.}\ \bibnamefont
  {Hopkins}}\ and\ \bibinfo {author} {\bibfnamefont {P.~M.}\ \bibnamefont
  {Norris}},\ }\href {\doibase 10.1115/1.2995623} {\bibfield  {journal}
  {\bibinfo  {journal} {Journal of Heat Transfer}\ }\textbf {\bibinfo {volume}
  {131}} (\bibinfo {year} {2009}),\ 10.1115/1.2995623}\BibitemShut {NoStop}%
\bibitem [{\citenamefont {Wang}\ \emph {et~al.}(2006)\citenamefont {Wang},
  \citenamefont {Wang},\ and\ \citenamefont {Zeng}}]{Wang2006}%
  \BibitemOpen
  \bibfield  {author} {\bibinfo {author} {\bibfnamefont {J.-S.}\ \bibnamefont
  {Wang}}, \bibinfo {author} {\bibfnamefont {J.}~\bibnamefont {Wang}}, \ and\
  \bibinfo {author} {\bibfnamefont {N.}~\bibnamefont {Zeng}},\ }\href {\doibase
  10.1103/PhysRevB.74.033408} {\bibfield  {journal} {\bibinfo  {journal} {Phys.
  Rev. B}\ }\textbf {\bibinfo {volume} {74}},\ \bibinfo {pages} {033408}
  (\bibinfo {year} {2006})}\BibitemShut {NoStop}%
\bibitem [{\citenamefont {Wang}\ \emph {et~al.}(2008)\citenamefont {Wang},
  \citenamefont {Wang},\ and\ \citenamefont {L{\"{u}}}}]{Wang2008}%
  \BibitemOpen
  \bibfield  {author} {\bibinfo {author} {\bibfnamefont {J.-S.}\ \bibnamefont
  {Wang}}, \bibinfo {author} {\bibfnamefont {J.}~\bibnamefont {Wang}}, \ and\
  \bibinfo {author} {\bibfnamefont {J.~T.}\ \bibnamefont {L{\"{u}}}},\ }\href
  {\doibase 10.1140/epjb/e2008-00195-8} {\bibfield  {journal} {\bibinfo
  {journal} {The European Physical Journal B}\ }\textbf {\bibinfo {volume}
  {62}},\ \bibinfo {pages} {381} (\bibinfo {year} {2008})}\BibitemShut
  {NoStop}%
\bibitem [{\citenamefont {Jiang}\ \emph {et~al.}(2009)\citenamefont {Jiang},
  \citenamefont {Wang},\ and\ \citenamefont {Li}}]{Jiang2009}%
  \BibitemOpen
  \bibfield  {author} {\bibinfo {author} {\bibfnamefont {J.-W.}\ \bibnamefont
  {Jiang}}, \bibinfo {author} {\bibfnamefont {J.-S.}\ \bibnamefont {Wang}}, \
  and\ \bibinfo {author} {\bibfnamefont {B.}~\bibnamefont {Li}},\ }\href
  {\doibase 10.1103/PhysRevB.79.205418} {\bibfield  {journal} {\bibinfo
  {journal} {Phys. Rev. B}\ }\textbf {\bibinfo {volume} {79}},\ \bibinfo
  {pages} {205418} (\bibinfo {year} {2009})}\BibitemShut {NoStop}%
\bibitem [{\citenamefont {Ouyang}\ \emph {et~al.}(2010)\citenamefont {Ouyang},
  \citenamefont {Chen}, \citenamefont {Xie}, \citenamefont {Yang},
  \citenamefont {Bao},\ and\ \citenamefont {Zhong}}]{Ouyang2010}%
  \BibitemOpen
  \bibfield  {author} {\bibinfo {author} {\bibfnamefont {T.}~\bibnamefont
  {Ouyang}}, \bibinfo {author} {\bibfnamefont {Y.}~\bibnamefont {Chen}},
  \bibinfo {author} {\bibfnamefont {Y.}~\bibnamefont {Xie}}, \bibinfo {author}
  {\bibfnamefont {K.}~\bibnamefont {Yang}}, \bibinfo {author} {\bibfnamefont
  {Z.}~\bibnamefont {Bao}}, \ and\ \bibinfo {author} {\bibfnamefont
  {J.}~\bibnamefont {Zhong}},\ }\href {\doibase 10.1088/0957-4484/21/24/245701}
  {\bibfield  {journal} {\bibinfo  {journal} {Nanotechnology}\ }\textbf
  {\bibinfo {volume} {21}},\ \bibinfo {pages} {245701} (\bibinfo {year}
  {2010})}\BibitemShut {NoStop}%
\bibitem [{\citenamefont {Zhou}\ \emph {et~al.}(2016)\citenamefont {Zhou},
  \citenamefont {Cai}, \citenamefont {Zhang},\ and\ \citenamefont
  {Zhang}}]{Zhou2016}%
  \BibitemOpen
  \bibfield  {author} {\bibinfo {author} {\bibfnamefont {H.}~\bibnamefont
  {Zhou}}, \bibinfo {author} {\bibfnamefont {Y.}~\bibnamefont {Cai}}, \bibinfo
  {author} {\bibfnamefont {G.}~\bibnamefont {Zhang}}, \ and\ \bibinfo {author}
  {\bibfnamefont {Y.-W.}\ \bibnamefont {Zhang}},\ }\href {\doibase
  10.1103/PhysRevB.94.045423} {\bibfield  {journal} {\bibinfo  {journal} {Phys.
  Rev. B}\ }\textbf {\bibinfo {volume} {94}},\ \bibinfo {pages} {045423}
  (\bibinfo {year} {2016})}\BibitemShut {NoStop}%
\bibitem [{\citenamefont {Ju}\ \emph {et~al.}(2017)\citenamefont {Ju},
  \citenamefont {Shiga}, \citenamefont {Feng}, \citenamefont {Hou},
  \citenamefont {Tsuda},\ and\ \citenamefont {Shiomi}}]{Ju2017}%
  \BibitemOpen
  \bibfield  {author} {\bibinfo {author} {\bibfnamefont {S.}~\bibnamefont
  {Ju}}, \bibinfo {author} {\bibfnamefont {T.}~\bibnamefont {Shiga}}, \bibinfo
  {author} {\bibfnamefont {L.}~\bibnamefont {Feng}}, \bibinfo {author}
  {\bibfnamefont {Z.}~\bibnamefont {Hou}}, \bibinfo {author} {\bibfnamefont
  {K.}~\bibnamefont {Tsuda}}, \ and\ \bibinfo {author} {\bibfnamefont
  {J.}~\bibnamefont {Shiomi}},\ }\href {\doibase 10.1103/PhysRevX.7.021024}
  {\bibfield  {journal} {\bibinfo  {journal} {Phys. Rev. X}\ }\textbf {\bibinfo
  {volume} {7}},\ \bibinfo {pages} {021024} (\bibinfo {year}
  {2017})}\BibitemShut {NoStop}%
\bibitem [{\citenamefont {Zhou}\ \emph {et~al.}(2017)\citenamefont {Zhou},
  \citenamefont {Cai}, \citenamefont {Zhang},\ and\ \citenamefont
  {Zhang}}]{Zhou2017}%
  \BibitemOpen
  \bibfield  {author} {\bibinfo {author} {\bibfnamefont {H.}~\bibnamefont
  {Zhou}}, \bibinfo {author} {\bibfnamefont {Y.}~\bibnamefont {Cai}}, \bibinfo
  {author} {\bibfnamefont {G.}~\bibnamefont {Zhang}}, \ and\ \bibinfo {author}
  {\bibfnamefont {Y.-W.}\ \bibnamefont {Zhang}},\ }\href {\doibase
  10.1038/s41699-017-0018-2} {\bibfield  {journal} {\bibinfo  {journal} {npj 2D
  Materials and Applications}\ }\textbf {\bibinfo {volume} {1}},\ \bibinfo
  {pages} {14} (\bibinfo {year} {2017})}\BibitemShut {NoStop}%
\bibitem [{\citenamefont {Zhou}\ \emph {et~al.}(2018)\citenamefont {Zhou},
  \citenamefont {Cai}, \citenamefont {Zhang},\ and\ \citenamefont
  {Zhang}}]{Zhou2018}%
  \BibitemOpen
  \bibfield  {author} {\bibinfo {author} {\bibfnamefont {H.}~\bibnamefont
  {Zhou}}, \bibinfo {author} {\bibfnamefont {Y.}~\bibnamefont {Cai}}, \bibinfo
  {author} {\bibfnamefont {G.}~\bibnamefont {Zhang}}, \ and\ \bibinfo {author}
  {\bibfnamefont {Y.-W.}\ \bibnamefont {Zhang}},\ }\href {\doibase
  10.1039/C7NR07779C} {\bibfield  {journal} {\bibinfo  {journal} {Nanoscale}\
  }\textbf {\bibinfo {volume} {10}},\ \bibinfo {pages} {480} (\bibinfo {year}
  {2018})}\BibitemShut {NoStop}%
\bibitem [{\citenamefont {Xu}\ \emph {et~al.}(2008)\citenamefont {Xu},
  \citenamefont {Wang}, \citenamefont {Duan}, \citenamefont {Gu},\ and\
  \citenamefont {Li}}]{Xu2008}%
  \BibitemOpen
  \bibfield  {author} {\bibinfo {author} {\bibfnamefont {Y.}~\bibnamefont
  {Xu}}, \bibinfo {author} {\bibfnamefont {J.-S.}\ \bibnamefont {Wang}},
  \bibinfo {author} {\bibfnamefont {W.}~\bibnamefont {Duan}}, \bibinfo {author}
  {\bibfnamefont {B.-L.}\ \bibnamefont {Gu}}, \ and\ \bibinfo {author}
  {\bibfnamefont {B.}~\bibnamefont {Li}},\ }\href {\doibase
  10.1103/PhysRevB.78.224303} {\bibfield  {journal} {\bibinfo  {journal} {Phys.
  Rev. B}\ }\textbf {\bibinfo {volume} {78}},\ \bibinfo {pages} {224303}
  (\bibinfo {year} {2008})}\BibitemShut {NoStop}%
\bibitem [{\citenamefont {Feng}\ and\ \citenamefont {Ruan}(2016)}]{Feng2016}%
  \BibitemOpen
  \bibfield  {author} {\bibinfo {author} {\bibfnamefont {T.}~\bibnamefont
  {Feng}}\ and\ \bibinfo {author} {\bibfnamefont {X.}~\bibnamefont {Ruan}},\
  }\href {\doibase 10.1103/PhysRevB.93.045202} {\bibfield  {journal} {\bibinfo
  {journal} {Phys. Rev. B}\ }\textbf {\bibinfo {volume} {93}},\ \bibinfo
  {pages} {045202} (\bibinfo {year} {2016})}\BibitemShut {NoStop}%
\bibitem [{\citenamefont {Feng}\ \emph {et~al.}(2017)\citenamefont {Feng},
  \citenamefont {Lindsay},\ and\ \citenamefont {Ruan}}]{Feng2017}%
  \BibitemOpen
  \bibfield  {author} {\bibinfo {author} {\bibfnamefont {T.}~\bibnamefont
  {Feng}}, \bibinfo {author} {\bibfnamefont {L.}~\bibnamefont {Lindsay}}, \
  and\ \bibinfo {author} {\bibfnamefont {X.}~\bibnamefont {Ruan}},\ }\href
  {\doibase 10.1103/PhysRevB.96.161201} {\bibfield  {journal} {\bibinfo
  {journal} {Phys. Rev. B}\ }\textbf {\bibinfo {volume} {96}},\ \bibinfo
  {pages} {161201} (\bibinfo {year} {2017})}\BibitemShut {NoStop}%
\bibitem [{\citenamefont {Ravichandran}\ and\ \citenamefont
  {Broido}(2020)}]{Ravichandran2020}%
  \BibitemOpen
  \bibfield  {author} {\bibinfo {author} {\bibfnamefont {N.~K.}\ \bibnamefont
  {Ravichandran}}\ and\ \bibinfo {author} {\bibfnamefont {D.}~\bibnamefont
  {Broido}},\ }\href {\doibase 10.1103/PhysRevX.10.021063} {\bibfield
  {journal} {\bibinfo  {journal} {Phys. Rev. X}\ }\textbf {\bibinfo {volume}
  {10}},\ \bibinfo {pages} {021063} (\bibinfo {year} {2020})}\BibitemShut
  {NoStop}%
\bibitem [{\citenamefont {Tielrooij}\ \emph {et~al.}(2018)\citenamefont
  {Tielrooij}, \citenamefont {Hesp}, \citenamefont {Principi}, \citenamefont
  {Lundeberg}, \citenamefont {Pogna}, \citenamefont {Banszerus}, \citenamefont
  {Mics}, \citenamefont {Massicotte}, \citenamefont {Schmidt}, \citenamefont
  {Davydovskaya}, \citenamefont {Purdie}, \citenamefont {Goykhman},
  \citenamefont {Soavi}, \citenamefont {Lombardo}, \citenamefont {Watanabe},
  \citenamefont {Taniguchi}, \citenamefont {Bonn}, \citenamefont
  {Turchinovich}, \citenamefont {Stampfer}, \citenamefont {Ferrari},
  \citenamefont {Cerullo}, \citenamefont {Polini},\ and\ \citenamefont
  {Koppens}}]{Tielrooij2018}%
  \BibitemOpen
  \bibfield  {author} {\bibinfo {author} {\bibfnamefont {K.-J.}\ \bibnamefont
  {Tielrooij}}, \bibinfo {author} {\bibfnamefont {N.~C.~H.}\ \bibnamefont
  {Hesp}}, \bibinfo {author} {\bibfnamefont {A.}~\bibnamefont {Principi}},
  \bibinfo {author} {\bibfnamefont {M.~B.}\ \bibnamefont {Lundeberg}}, \bibinfo
  {author} {\bibfnamefont {E.~A.~A.}\ \bibnamefont {Pogna}}, \bibinfo {author}
  {\bibfnamefont {L.}~\bibnamefont {Banszerus}}, \bibinfo {author}
  {\bibfnamefont {Z.}~\bibnamefont {Mics}}, \bibinfo {author} {\bibfnamefont
  {M.}~\bibnamefont {Massicotte}}, \bibinfo {author} {\bibfnamefont
  {P.}~\bibnamefont {Schmidt}}, \bibinfo {author} {\bibfnamefont
  {D.}~\bibnamefont {Davydovskaya}}, \bibinfo {author} {\bibfnamefont {D.~G.}\
  \bibnamefont {Purdie}}, \bibinfo {author} {\bibfnamefont {I.}~\bibnamefont
  {Goykhman}}, \bibinfo {author} {\bibfnamefont {G.}~\bibnamefont {Soavi}},
  \bibinfo {author} {\bibfnamefont {A.}~\bibnamefont {Lombardo}}, \bibinfo
  {author} {\bibfnamefont {K.}~\bibnamefont {Watanabe}}, \bibinfo {author}
  {\bibfnamefont {T.}~\bibnamefont {Taniguchi}}, \bibinfo {author}
  {\bibfnamefont {M.}~\bibnamefont {Bonn}}, \bibinfo {author} {\bibfnamefont
  {D.}~\bibnamefont {Turchinovich}}, \bibinfo {author} {\bibfnamefont
  {C.}~\bibnamefont {Stampfer}}, \bibinfo {author} {\bibfnamefont {A.~C.}\
  \bibnamefont {Ferrari}}, \bibinfo {author} {\bibfnamefont {G.}~\bibnamefont
  {Cerullo}}, \bibinfo {author} {\bibfnamefont {M.}~\bibnamefont {Polini}}, \
  and\ \bibinfo {author} {\bibfnamefont {F.~H.~L.}\ \bibnamefont {Koppens}},\
  }\href {\doibase 10.1038/s41565-017-0008-8} {\bibfield  {journal} {\bibinfo
  {journal} {Nature Nanotechnology}\ }\textbf {\bibinfo {volume} {13}},\
  \bibinfo {pages} {41} (\bibinfo {year} {2018})}\BibitemShut {NoStop}%
\bibitem [{\citenamefont {Alborzi}\ \emph {et~al.}(2020)\citenamefont
  {Alborzi}, \citenamefont {Rajabpour},\ and\ \citenamefont
  {Montazeri}}]{Alborzi2020}%
  \BibitemOpen
  \bibfield  {author} {\bibinfo {author} {\bibfnamefont {M.~S.}\ \bibnamefont
  {Alborzi}}, \bibinfo {author} {\bibfnamefont {A.}~\bibnamefont {Rajabpour}},
  \ and\ \bibinfo {author} {\bibfnamefont {A.}~\bibnamefont {Montazeri}},\
  }\href {\doibase https://doi.org/10.1016/j.ijthermalsci.2019.106237}
  {\enquote {\bibinfo {title} {Heat transport in 2d van der waals
  heterostructures: An analytical modeling approach},}\ } (\bibinfo {year}
  {2020})\BibitemShut {NoStop}%
\bibitem [{\citenamefont {Zhou}\ \emph {et~al.}(2020)\citenamefont {Zhou},
  \citenamefont {Zhang}, \citenamefont {Wang},\ and\ \citenamefont
  {Zhang}}]{Zhou2020}%
  \BibitemOpen
  \bibfield  {author} {\bibinfo {author} {\bibfnamefont {H.}~\bibnamefont
  {Zhou}}, \bibinfo {author} {\bibfnamefont {G.}~\bibnamefont {Zhang}},
  \bibinfo {author} {\bibfnamefont {J.-S.}\ \bibnamefont {Wang}}, \ and\
  \bibinfo {author} {\bibfnamefont {Y.-W.}\ \bibnamefont {Zhang}},\ }\href
  {\doibase 10.1103/PhysRevB.101.235305} {\bibfield  {journal} {\bibinfo
  {journal} {Phys. Rev. B}\ }\textbf {\bibinfo {volume} {101}},\ \bibinfo
  {pages} {235305} (\bibinfo {year} {2020})}\BibitemShut {NoStop}%
\bibitem [{\citenamefont {Li}\ \emph {et~al.}(2012{\natexlab{b}})\citenamefont
  {Li}, \citenamefont {Agarwalla},\ and\ \citenamefont {Wang}}]{Li2012a}%
  \BibitemOpen
  \bibfield  {author} {\bibinfo {author} {\bibfnamefont {H.}~\bibnamefont
  {Li}}, \bibinfo {author} {\bibfnamefont {B.~K.}\ \bibnamefont {Agarwalla}}, \
  and\ \bibinfo {author} {\bibfnamefont {J.-S.}\ \bibnamefont {Wang}},\ }\href
  {\doibase 10.1103/PhysRevE.86.011141} {\bibfield  {journal} {\bibinfo
  {journal} {Phys. Rev. E}\ }\textbf {\bibinfo {volume} {86}},\ \bibinfo
  {pages} {011141} (\bibinfo {year} {2012}{\natexlab{b}})}\BibitemShut
  {NoStop}%
\bibitem [{\citenamefont {Balandin}(2020)}]{Balandin2020}%
  \BibitemOpen
  \bibfield  {author} {\bibinfo {author} {\bibfnamefont {A.~A.}\ \bibnamefont
  {Balandin}},\ }\href {\doibase 10.1021/acsnano.0c02718} {\bibfield  {journal}
  {\bibinfo  {journal} {ACS Nano}\ }\textbf {\bibinfo {volume} {14}},\ \bibinfo
  {pages} {5170} (\bibinfo {year} {2020})}\BibitemShut {NoStop}%
\end{thebibliography}%

\end{document}